\documentclass[sigconf]{acmart}

\AtBeginDocument{%
  }

\begin{document}

\title{Analyzing the factors that are involved in length of inpatient stay at the hospital for diabetes patients}

\author{Jorden Lam}
\affiliation{%
  \institution{Harrisburg University of Science and Technology}
  \city{Harrisburg}
  \state{Pennsylvania}
  \country{USA}
}

\author{Kunpeng Xu}
\affiliation{%
  \institution{Université de Sherbrooke}
  \city{Sherbrooke}
  \state{Quebec}
  \country{Canada}
}


\begin{abstract}
The paper investigates the escalating concerns surrounding the surge in diabetes cases, exacerbated by the COVID-19 pandemic, and the subsequent strain on medical resources. The research aims to construct a predictive model quantifying factors influencing inpatient hospital stay durations for diabetes patients, offering insights to hospital administrators for improved patient management strategies. The literature review highlights the increasing prevalence of diabetes, emphasizing the need for continued attention and analysis of urban-rural disparities in healthcare access. International studies underscore the financial implications and healthcare burden associated with diabetes-related hospitalizations and complications, emphasizing the significance of effective management strategies. The methodology involves a quantitative approach, utilizing a dataset comprising 10,000 observations of diabetic inpatient encounters in U.S. hospitals from 1999 to 2008. Predictive modeling techniques, particularly Generalized Linear Models (GLM), are employed to develop a model predicting hospital stay durations based on patient demographics, admission types, medical history, and treatment regimen. The results highlight the influence of age, medical history, and treatment regimen on hospital stay durations for diabetes patients. Despite model limitations, such as heteroscedasticity and deviations from normality in residual analysis, the findings offer valuable insights for hospital administrators in patient management. The paper concludes with recommendations for future research to address model limitations and explore the implications of predictive models on healthcare management strategies, ensuring equitable patient care and resource allocation.
\end{abstract}


\keywords{Diabetes, GLM}

\maketitle

\section{Introduction}
In recent years, the surge in diabetes cases has raised significant concerns, intensifying the demand for medical resources, \cite{wang2021trends} and \cite{fang2021trends,xu2023drnet}. This is a crucial endeavor amidst the mounting challenges posed by the increasing diabetic population and the lingering effects of the COVID-19 pandemic. Current research aims to construct a model that quantifies the factors influencing inpatient hospital stay durations for diabetes patients, enabling hospital administrators to gain a deeper understanding of patient needs and enhance their management strategies.

\section{Literature Review}\label{literature-review}

Before the COVID era, the pattern of increasing diabetes patients had been a wake-up call due to the substantial consumption of medical resources. Both \cite{wang2021trends} and \cite{fang2021trends} analyzed U.S. data from 1999 to 2018, highlighting the rising prevalence of diabetes, which necessitates ongoing attention. According to \cite{mercado2021differences}, a comparison between rural and urban data from the same period revealed that rural areas face worse conditions than their urban counterparts. These findings align with the global perspective provided by \cite{bach2014high} in Melbourne, where the high prevalence of diabetes among inpatients significantly impacts healthcare expenditure.

Diabetes is not confined to the elderly. \cite{icks2001hospitalization} compared hospitalization rates and costs between diabetic children and adolescents (aged 1–19 years) in Germany and the general population, finding that diabetic children and adolescents had approximately three times higher hospitalization risk and three times more hospital days than their age-matched counterparts. This underscores the need for targeted healthcare strategies across different age groups. Similarly, \cite{bala2022length} conducted a multivariate regression analysis in a Romanian public hospital, revealing that higher adjusted average costs per hospitalization episode and longer hospital stays were linked to increasing age, the presence of cardiovascular diseases, chronic kidney disease, and foot ulcerations. \cite{cheng2019costs} also investigated the costs and length of hospitalizations related to diabetes complications in Taiwan, highlighting the global burden of the disease.

Complications related to diabetes warrant significant attention. \cite{clarke2003impact} developed a model to estimate immediate and long-term healthcare costs associated with various diabetes-related complications in Type 2 diabetes patients. Hypoglycemia, as noted by \cite{turchin2009hypoglycemia} and \cite{brodovicz2013association}, significantly extends the length of stay, highlighting the critical need for effective management of such complications. \cite{korbel2015diabetes} emphasized the frequent need for emergency department care and hospitalization for infection management among diabetic patients, underlining the considerable socioeconomic burden. This aligns with global studies such as \cite{xu2022data}, who utilized kernel-based methods to uncover nonlinear relationships in healthcare data, demonstrating the importance of advanced analytical techniques in managing diabetes complications.

Predictive modeling plays a crucial role in managing diabetes-related hospitalizations. \cite{alturki2019predictors} addressed the growing concern of unplanned hospital readmissions and extended hospital stays, suggesting that early prediction of these outcomes could facilitate targeted interventions and improved resource allocation. \cite{koziol2021predictors} investigated risk factors associated with rehospitalization and mortality following diabetes-related hospital admissions in Poland, further illustrating the need for robust predictive models. Scholars have utilized various machine learning techniques to predict hospital stays. \cite{barsasella2022machine} demonstrated that linear regression works well in Length of Stay (LoS) prediction. Advanced techniques, such as the stacked ensemble method discussed by \cite{alahmar2018application}, highlight the potential for data-driven strategies to improve healthcare management and outcomes for diabetes patients. Recent advancements, such as the work by \cite{xu2024kernel,xu2025towards}, on time series analysis provide deeper insights into the temporal aspects of patient data.

Over the past three years, several COVID-19 outbreaks have caused varying degrees of bed shortages, especially ICU beds, in both developed countries with ample medical resources and developing countries with already poor medical conditions \cite{sen2021closer,xu2025drift2matrix}. Sometimes, the COVID-19 testing process itself, rather than the infection, delays patient discharge due to requirements for negative test results before patients can be transferred to nursing care facilities \cite{olanipekun2021impact}. This often results in an additional 2-3 days of hospital stay before discharge. The pandemic has also underscored the importance of advanced modeling techniques. \cite{xu2022multi,xu2020kernel,chen2021self} discussed the application of multi-view kernel clustering to explore the nonlinear relationships in healthcare data, which is crucial for understanding the complex interactions between COVID-19 and diabetes management.

Recent advancements in kernel-based methods offer promising tools for quantifying factors affecting hospital stay durations for diabetes patients by uncovering nonlinear relationships \cite{xu2024rhine,pethunachiyar2020classification,xu2018self}. The latest model, Kolmogorov-Arnold Networks (KAN), introduces a deep learning framework with high interpretability, offering a promising strategy for analyzing complex, time-dependent healthcare data \cite{xu2024kolmogorov,chen2022clustering,xu2024kan4drift,chen2022dynamic,xu2024kan}. This concept of interpretability is similar to that of \cite{nwegbu2022novel,xu2024rhine,xu2024wormhole}, who emphasized the importance of dynamic regime discovery in forecasting states of data (e.g., healthcare data), and offers new ways to improve the predictive accuracy of diabetes management.

In summary, factors such as age, sex, health condition, and treatment utilization are strongly related to the length of stay in hospitals. Regression techniques have proven effective in exploring these relationships. Considering the immense pressure from the increasing diabetic population and the lasting effects of COVID-19, developing a model that provides insights into the determinants of inpatient hospital stay durations is crucial. Such a model would enable hospital administrators to better understand and manage patient needs, ultimately enhancing healthcare outcomes.

\section{Methods}
The research methodology adopted for this study is primarily quantitative and involves predictive modeling. The objective is to develop a model that can effectively predict the length of inpatient hospital stays for patients with diabetes by using Regression. This model will serve as a valuable tool for hospital administrators in understanding and managing the needs of diabetic patients more efficiently.

\subsection{Participants}
The dataset used for this research consists of 10,000 observations, representing historical inpatient encounters of diabetic patients in U.S. hospitals spanning the period between 1999 and 2008. Each observation encompasses comprehensive information about the specific hospital stay, details about the patient, their recent medical treatments, and the treatments administered upon admission.

The research approach involves constructing a predictive model that can identify and leverage key factors and variables within the dataset to make accurate predictions regarding the length of inpatient visits for diabetic patients. It's important to note that this predictive model is tailored specifically to diabetic patients and may not be directly applicable to non-diabetic patients. The null hypothesis assumes length of stay has no relationship with age, sex, health condition or utilization of treatment.
  \(H_1\): There is a significant positive/negative relationship between at least one independent variables among the dataset provided and the length of the hospital stay.
  \(H_0\): There is no significant relationship between any variable and the length of the hospital stay.

The dataset encompasses a range of variables that are instrumental in predicting the length of hospital stays for diabetic patients. These variables include the length of the hospital stay in days, which serves as the primary outcome variable. Additionally, demographic information of the patients was considered, providing insights into the age, gender, and other relevant characteristics of the study population, as independent variables.

The type of hospital admission, a critical factor influencing the length of stay, was also included in the dataset. This variable helps categorize admissions as elective, urgent, or emergency, shedding light on the urgency and nature of the medical condition. Furthermore, the dataset incorporated a history of medical activity in the 12 months preceding the hospital stay. This historical perspective is invaluable in understanding the patient's medical background and its impact on the current hospitalization.

Lastly, specific information pertaining to changes in diabetes medication upon admission was included, particularly focusing on metformin and insulin. These variables hold significance in assessing the treatment regimen and its influence on the length of hospital stays.
\subsection{Procedure}
The Procedures section outlines the methodology employed for the inclusion of units in the study, specifically focusing on the collection of the dataset and its relevant details. The dataset utilized for this study is derived from the "Diabetes 130-US hospitals for years 1999-2008," which is publicly available and maintained by the UCI Machine Learning Repository \footnote{\url{https://archive.ics.uci.edu/dataset/296/diabetes+130-us+hospitals+for+years+1999-2008}}.

The dataset comprises information from 130 U.S. hospitals and includes patient records from the years 1999 to 2008. This dataset is a result of a comprehensive data compilation process from various healthcare institutions.

The dataset was constructed by gathering patient records related to diabetes-related hospital admissions. To be included in the dataset, the hospital encounters needed to satisfy specific criteria, ensuring that they were relevant to diabetes patients. It involved assembling relevant data fields, including patient demographics, hospital stay details, admission types, prior medical history, and medication changes upon admission.

The dataset is publicly accessible through the UCI Machine Learning Repository, making it readily available for research purposes. Researchers and analysts can access and download the dataset to conduct studies and build predictive models. By the way, ethical considerations were paramount during data collection. Personal identifiers were removed or anonymized to protect patient privacy and comply with data privacy regulations.

\subsection{Measures}
This section provides a comprehensive overview of the research methodology employed in this study, emphasizing the measures, metrics, and modeling techniques used. It underscores the reliability and validity of these tools while highlighting the rationale behind their selection. The primary objective of this research is to predict the length of inpatient hospital stays for diabetes patients based on historical data. The study's methodology involves rigorous model development, validation, and evaluation to identify factors influencing hospitalization duration accurately.

The primary outcome metric, the length of inpatient hospital stays in days, serves as the dependent variable in this study. This metric is widely used in healthcare research to assess patient care outcomes and resource utilization. It provides a quantitative measure of the duration of hospitalization for diabetes patients.

To assess the performance of various predictive models, several standard metrics were employed, including:

*Mean Absolute Error (MAE): MAE measures the average absolute difference between the predicted and actual lengths of hospital stays. It is a reliable indicator of model accuracy.
Root Mean Square Error (RMSE): RMSE quantifies the square root of the average squared differences between predicted and actual values, providing insights into the model's predictive accuracy.

*R-squared (R2): R2 measures the proportion of the variance in the dependent variable (length of stay) explained by the independent variables. It assesses the goodness of fit of the model.

*Akaike Information Criterion (AIC): AIC is a measure of the relative quality of a statistical model. It balances the model's goodness of fit with its complexity, penalizing models that are too complex. Lower AIC values indicate better-fitting models.

*Bayesian Information Criterion (BIC): BIC is similar to AIC but places a stronger penalty on complex models. It aims to identify the simplest yet most effective model. Lower BIC values indicate better model selection.

The research employed a rigorous approach to model development and evaluation. The model was trained using 70

The selected model for this study is the Generalized Linear Model (GLM). GLM is a well-established statistical technique for modeling the relationship between a dependent variable and one or more independent variables. It is widely used in healthcare research due to its interpretability and performance in explaining factors affecting patient outcomes.

Regarding validity and reliability, the metrics and tools used in this study have demonstrated their validity and reliability in various healthcare research contexts. Length of inpatient stay is a standard and widely accepted measure of hospitalization duration. Model evaluation metrics like MAE, RMSE, R2, AIC, and BIC are well-established and extensively used in predictive modeling.

The selection of the GLM model is based on its robustness and interpretability. GLM has been successfully applied in healthcare research to analyze factors affecting hospital stays and patient outcomes. The rigorous model calibration and validation process further enhances the trustworthiness of the selected model's performance.
  
\subsection{Analysis}
The analysis phase of this study involved a systematic approach to developing, evaluating, and selecting the most suitable predictive model for determining the length of inpatient hospital stays for diabetes patients. Several key steps were undertaken to ensure the robustness and reliability of our predictive model.
\subsubsection{Exploratory Data Analysis and Feature Engineering}
Initially, we delve into the dataset to understand its characteristics and gain initial insights. EDA that will be done includes calculating descriptive statistics, creating data visualization, exploring correlation analysis and accessing data distribution. Then, we will conduct feature engineering to prepare the dataset for modeling. This step involves modifying specific features to enhance their compatibility with various modeling techniques.

Here are some more details of processing data.

There were 3 unknown/invalid values for the gender variable. Because there were only 3 records out of 10,000 total records, these rows were removed from the data.

There were 9,691 records with missing values for the weight variable. Because most of the weight data was missing, the variable was removed from the dataset.

The admin\_type\_id variable was coded as a numeric variable. Since the numeric values are codes representing categorical data, the variable was changed to a factor variable. There were 1,021 records where the admission type was unavailable, which could be viewed as missing. We do not know if these are missing because of a data collection error or the admission type is routinely unavailable. Because we do not know whether it is missing at random, we should keep it and see whether it being unavailable has predictive power.

The race variable contained 226 missing values. Similar to the admin\_type\_id missing data, we do not know if these are missing because of a data collection error or the race is routinely unknown. Because we do not know whether it is missing at random, we should keep the variable and see whether missing race has predictive power. A new race category was created called ``Missing.'' I also combined the ``Asian'', ``Hispanic'', and ``Other'' levels because they each had somewhat low frequencies and similar relationships to the days variable.

\begin{verbatim}
## 
##         Missing AfricanAmerican           Other       Caucasian 
##             226            1848             392            7531
\end{verbatim}
The num\_meds variable had values ranging from 1 to 67. It seems unlikely that in a large dataset there would be no individuals that took 0 medications in the prior year. This is suspicious and should be investigated because it could be an indicator of invalid data, but the values look reasonable aside from that, so I will use the variable without alterations.

The factor variable levels were reordered so that the most frequent level was first. After the changes 9,997 records remained in the dataset. To explore the data, here exhibits some descriptive statistics (AppendixA).

The target variable is the number of days between admission into and discharge from the hospital.
The variable takes on integer values from 1 to 14. The center of the distribution is around 4 to 4.5
days based on the median/mean. From the bar chart below, we can see that the distribution is
skewed right with 2-3 days being the most frequent length of stay in the hospital.(AppendixB)

The insulin variable indicates whether, upon admission, insulin was prescribed or there was a
change in the dosage. About half of the records did not have insulin prescribed upon admission
and these records were admitted on average over a day less than records where insulin was
increased upon admission. The boxplots below show that the median and 3rd quartile number of
days are also lower when insulin is not prescribed. Changes to insulin dosages also had higher
mean days. I selected this variable because (1) each variable level has over 1000 records and a
noticeable difference in mean days and (2) It makes intuitive sense that requiring a medication or
change to it upon arrival might lead to a need to monitor a patient over a period of time, increasing
the length of stay.(AppendixC)

The age variable is a factor variable, and frequency counts can be seen in the bar chart and table
below. Most of our data has ages between 50 and 90, with 70-80 containing the most data and thus
it is the baseline. If age were numeric, we would say it was skewed left. The table and box plots
below show that for the most part as age increases, days increases. There is a sizeable difference
(over 1.5 days) between mean days for the age bins with the highest mean days and the lowest
mean days. I chose this variable because it makes sense that older patients tend to stay longer
since they tend to be in poorer health, and the relationship appears to be strong.(AppendixD)

The num\_meds variable takes on integer values from 1 to 67. The center of the distribution is
around 15 to 16 based on the median/mean. From the bar chart below, we can see that the
distribution is skewed right with 13 being the most frequent number of medications. The
correlation between num\_meds and days was 0.472, which was the strongest correlation among
the numeric variables. The scatterplot, with point size/color based on frequency below shows that
as the number of medications increases, the days increases. Like the other variables selected, this
relationship makes intuitive sense -- patients taking many medications likely have more underlying
conditions, have poorer overall health, and may require a longer hospital stay to address those
concerns. I selected this variable because the relationship to the target variable was the strongest
of the numeric variables and the narrative makes sense.(AppendixE)

Hence, Insulin, age, and num\_meds appear to have strong relationships to the target variable.
\subsubsection{Model Building}\label{model-building}

We will experiment with various modeling techniques to identify the optimal approach for explaining the factors influencing the length of hospital stays. Each model is meticulously calibrated using 70\% of the dataset, and its performance was rigorously assessed using the remaining 30\%. This process allows us to gauge the models' ability to capture data patterns and generalize effectively to new data.

\begin{verbatim}
## Warning: package 'caret' was built under R version 4.2.3
\end{verbatim}

\begin{verbatim}
## [1] "TRAIN"
\end{verbatim}

\begin{verbatim}
## [1] 4.397857
\end{verbatim}

\begin{verbatim}
## [1] "TEST"
\end{verbatim}

\begin{verbatim}
## [1] 4.435769
\end{verbatim}

\begin{verbatim}
## [1] "ALL"
\end{verbatim}

\begin{verbatim}
## [1] 4.409223
\end{verbatim}

Here is a GLM with Poisson distribution with a log link function (simple multiple linear regression is just a special case of GLM that is Normal distribution with an identical link), since Poisson is a good distribution to fit counts data and log link is easy to interpret and make sure the prediction is always positive. I don't try any other distribution or link, but just select by intuition. The output from the GLM with all variables selected is below, as well as the Pearson goodness-of-fit statistic on test data.(AppendixF)

\subsubsection{Feature Selection}\label{feature-selection}

Feature selection will be performed by using multiple metrics discussed in the previous Measures section, which tends to remove unnecessary variables in order to prevent overfitting.(AppendixG)

I then perform feature selection by using forward selection with BIC, which tends to remove more variables in order to prevent overfitting. As a result, the following variables were retained.(AppendixH)

\subsubsection{Model Validation}\label{model-validation}

Ultimately, a model who has superior performance in capturing data patterns and providing valuable insights into the factors influencing the length of hospital stays will be our predictive model of choice. We may also check its accuracy in-sample and out-sample as well as if the model holds the basic assumptions. Furthermore, to make it the ideal choice for our analysis, the selected model should also offer transparency and interpretability.

The Pearson goodness-of-fit statistic for both training and test data. They are pretty close, which means the the selected model still keeps much info as it can while preventing overfitting.

\begin{enumerate}
\def\labelenumi{\arabic{enumi}.}

\item
  Use full model on training data
\end{enumerate}

\begin{verbatim}
## [1] 1.525157
\end{verbatim}

\begin{enumerate}
\def\labelenumi{\arabic{enumi}.}
\setcounter{enumi}{1}

\item
  Use full model on test data
\end{enumerate}

\begin{verbatim}
## [1] 1.558029
\end{verbatim}

\begin{enumerate}
\def\labelenumi{\arabic{enumi}.}
\setcounter{enumi}{2}

\item
  Use selected model on test data
\end{enumerate}

\begin{verbatim}
## [1] 1.567946
\end{verbatim}

The following plot of residuals versus fitted values shows that the model performs is not very well. The dots are not centered near zero or spread symmetrically in each direction, indicating homogeneity or homoscedasticity is not hold.

The q‐q plot shows that the normal distribution assumption (for the residuals) is not maintained strictly. Points bend up and to the left of the line which means a right skew distribution.(AppendixI)

Thus, it appears a fatter‐tailed model may do better.

\section{Results}\label{results}

The model coefficients can be used to gain insights about the factors affecting a patient's length of stay. The model starts with a baseline predicted length of stay for each patient of 1.99 days. Then, it applies the factors above (AppendixH) based on the patient data. Note that values have been rounded. Multiplying by factors greater than 1 increase the predicted length of stay, while multiplying by factors less than 1 decrease the predicted length of stay. Due to the use of the log link, an appropriate way to interpret coefficients is to exponentiate them.

Many of the factors affecting the length of stay make intuitive sense. Starting at age 50, the length
of stay on average increases as age increases, which is not surprising as older patients tend to have declining health. Increased treatments (medications and diagnoses) in the prior year also led to longer stays on average.

\section{Discussion}\label{discussion}

\subsection{Limitations}\label{limitations}

While the research endeavors to construct a robust predictive model for determining inpatient hospital stay durations for diabetes patients, several limitations warrant consideration. Firstly, the model's performance, as indicated by residual analysis, reveals deviations from homogeneity and normality assumptions, suggesting potential shortcomings in capturing underlying data patterns. Addressing these limitations may require exploring alternative modeling techniques or incorporating additional variables to enhance model accuracy.

Furthermore, the dataset's temporal scope spans from 1999 to 2008, potentially limiting the model's applicability to contemporary healthcare settings. Updating the dataset with more recent patient encounters could provide a more comprehensive understanding of current trends and factors influencing hospital stay durations for diabetes patients.

\subsection{Future Work}\label{future-work}

Despite the identified limitations, the study lays the groundwork for future research avenues aimed at improving predictive modeling of hospital stay durations for diabetes patients. One potential avenue for future work involves refining the predictive model by exploring alternative modeling techniques or incorporating additional variables, such as fatter-tailed models, to enhance predictive accuracy.

Future research could focus on validating the predictive model's performance in diverse healthcare settings and patient populations to assess its generalizability and scalability. Collaborative efforts with healthcare institutions and stakeholders could facilitate the implementation of predictive models in clinical practice, enabling healthcare providers to optimize patient management strategies and resource allocation.

Furthermore, exploring the ethical implications and potential biases associated with predictive modeling in healthcare is essential to ensure equitable patient care and mitigate algorithmic biases. Future research could investigate the impact of predictive models on healthcare disparities and develop strategies to address potential biases and disparities in model implementation.

Overall, future research endeavors should prioritize addressing the identified limitations, refining predictive models, and fostering collaboration between researchers, healthcare practitioners, and stakeholders to advance the application of predictive analytics in improving patient outcomes and healthcare delivery.

\bibliographystyle{IEEEtran} 
\bibliography{sample-base}

\newpage
\appendix
\onecolumn
\begin{Large}
   \textbf{APPENDIX} 
\end{Large}
\section{Descriptive Statistics of All Variables} \label{App:AppendixA}

\vspace{0pt}
\begin{verbatim}
##       days           gender          age                    race     
##  Min.   : 1.000   Female:5338   [70-80):2541   Caucasian      :7531  
##  1st Qu.: 2.000   Male  :4659   [60-70):2228   Missing        : 226  
##  Median : 4.000                 [50-60):1726   AfricanAmerican:1848  
##  Mean   : 4.409                 [80-90):1676   Other          : 392  
##  3rd Qu.: 6.000                 [40-50): 931                         
##  Max.   :14.000                 [30-40): 372                         
##                                 (Other): 523                         
##  admit_type_id  metformin      insulin     readmitted   num_procs    
##  1:5289        No    :8024   No    :4742   NO :5370   Min.   :0.000  
##  2:1870        Down  :  60   Down  :1198   <30:1102   1st Qu.:0.000  
##  3:1817        Steady:1799   Steady:2928   >30:3525   Median :1.000  
##  4:1021        Up    : 114   Up    :1129              Mean   :1.345  
##                                                       3rd Qu.:2.000  
##                                                       Max.   :6.000  
##                                                                      
##     num_meds         num_ip          num_diags     
##  Min.   : 1.00   Min.   : 0.0000   Min.   : 1.000  
##  1st Qu.:10.00   1st Qu.: 0.0000   1st Qu.: 6.000  
##  Median :15.00   Median : 0.0000   Median : 8.000  
##  Mean   :16.16   Mean   : 0.6388   Mean   : 7.442  
##  3rd Qu.:20.00   3rd Qu.: 1.0000   3rd Qu.: 9.000  
##  Max.   :67.00   Max.   :21.0000   Max.   :16.000  
## 
\end{verbatim}

\section{Distribution of Target Variable - Length of Hospital Stay} \label{App:AppendixB}
\vspace{20pt}
\includegraphics[width=0.8\linewidth]{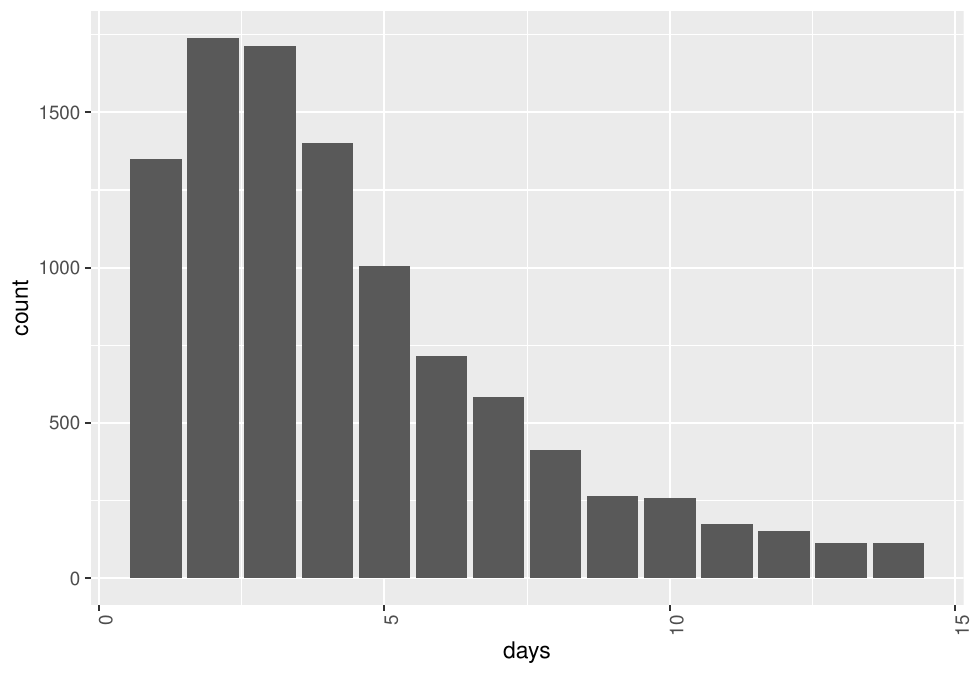}

\newpage
\section{Details of Independent Variable - Insulin} \label{App:AppendixC}

\begin{verbatim}
## Warning: `group_by_()` was deprecated in dplyr 0.7.0.
## i Please use `group_by()` instead.
## i See vignette('programming') for more help
## Call `lifecycle::last_lifecycle_warnings()` to see where this warning was
## generated.
\end{verbatim}

\begin{verbatim}
## # A tibble: 3 x 4
##   readmitted  mean median     n
##   <fct>      <dbl>  <dbl> <int>
## 1 NO          4.23      3  5370
## 2 <30         4.77      4  1102
## 3 >30         4.57      4  3525
\end{verbatim}
\vspace{30pt}
\includegraphics{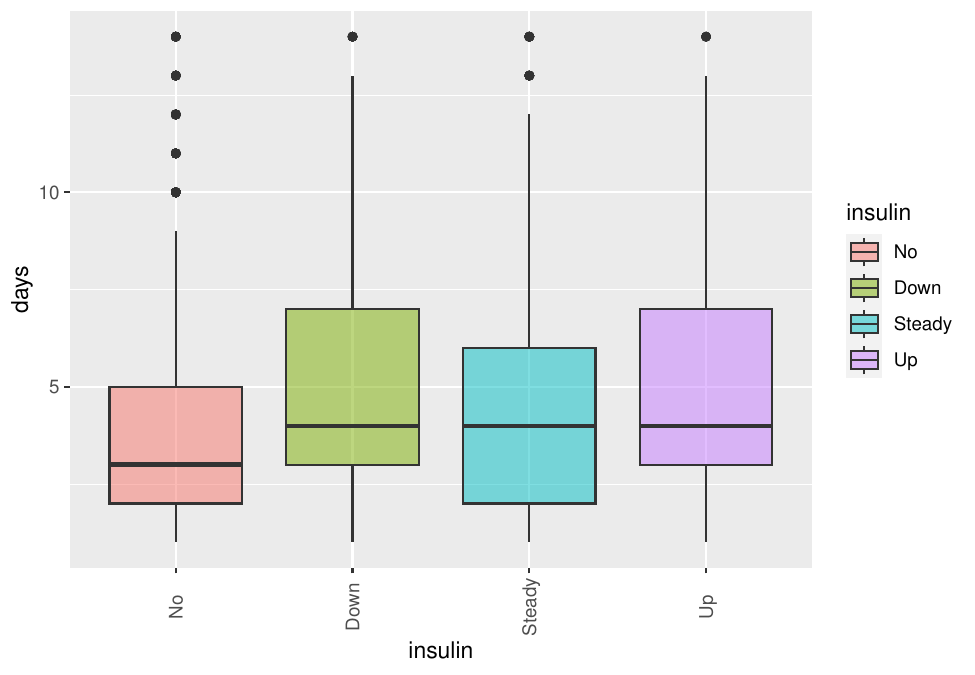}

\newpage
\section{Details of Independent Variable - Age} \label{App:AppendixD}
\vspace{30pt}
\includegraphics{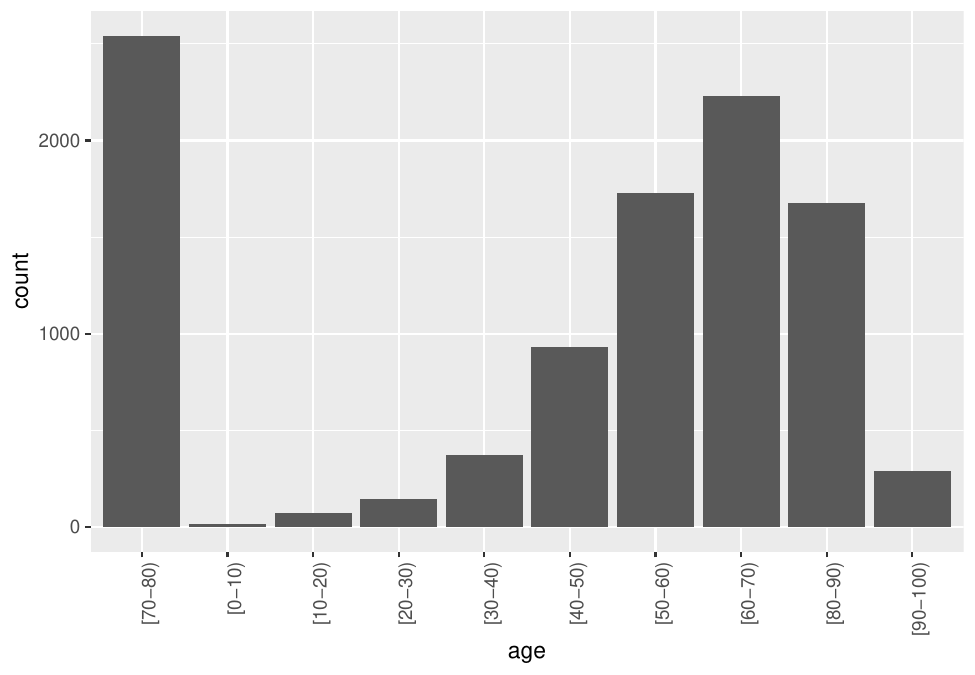}

\begin{verbatim}
## Warning: `group_by_()` was deprecated in dplyr 0.7.0.
## i Please use `group_by()` instead.
## i See vignette('programming') for more help
## Call `lifecycle::last_lifecycle_warnings()` to see where this warning was
## generated.
\end{verbatim}

\begin{verbatim}
## # A tibble: 10 x 4
##    age       mean median     n
##    <fct>    <dbl>  <dbl> <int>
##  1 [70-80)   4.66      4  2541
##  2 [0-10)    3.22      3    18
##  3 [10-20)   3.12      2    72
##  4 [20-30)   3.52      3   145
##  5 [30-40)   3.80      3   372
##  6 [40-50)   3.99      3   931
##  7 [50-60)   4.09      3  1726
##  8 [60-70)   4.41      4  2228
##  9 [80-90)   4.82      4  1676
## 10 [90-100)  4.74      4   288
\end{verbatim}

\newpage
\section{Details of Independent Variable - Number of Medication} \label{App:AppendixE}
\vspace{30pt}
\includegraphics[width=0.7\linewidth]{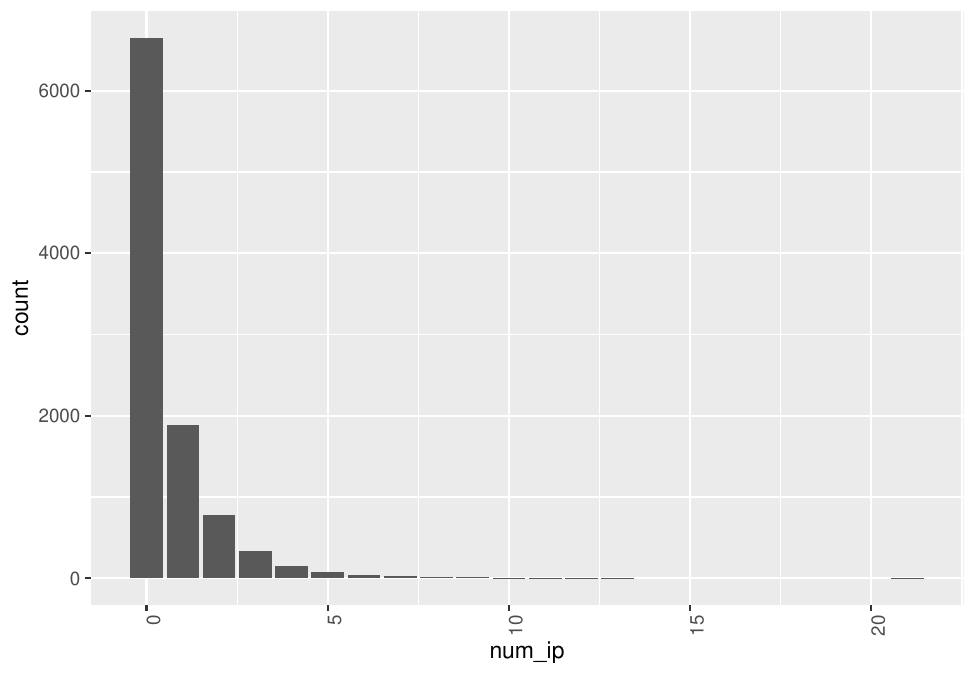}

\includegraphics[width=0.8\linewidth]{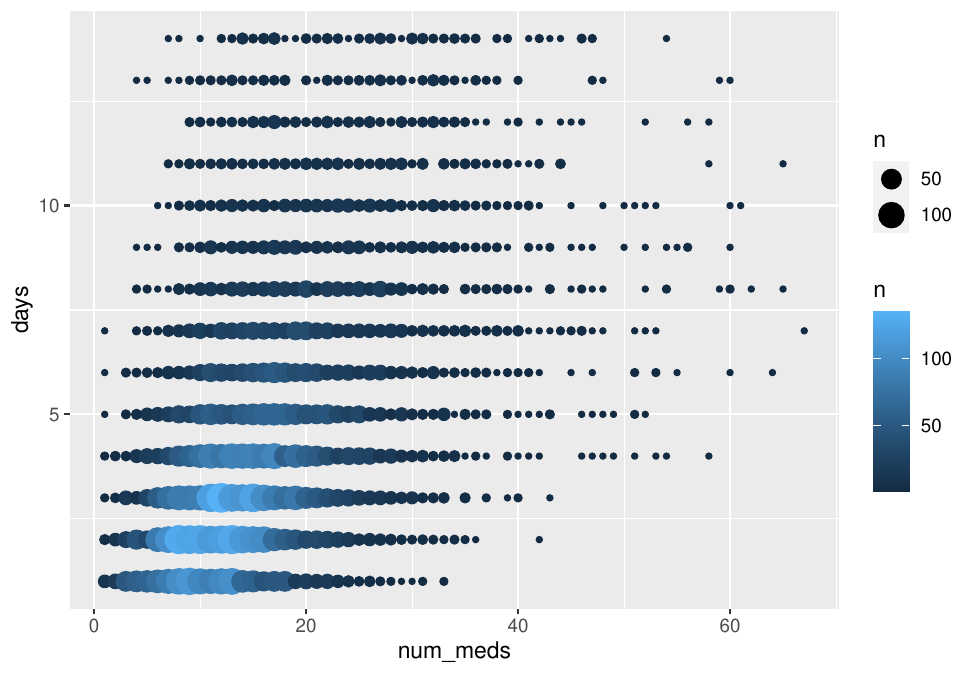}

\newpage
\section{GLM with Poisson distribution with a log link function} \label{App:AppendixF}

\begin{verbatim}
## 
## Call:
## glm(formula = days ~ ., family = poisson(link = "log"), data = data.train)
## 
## Deviance Residuals: 
##     Min       1Q   Median       3Q      Max  
## -3.6784  -0.9793  -0.2836   0.5744   4.8217  
## 
## Coefficients:
##                       Estimate Std. Error z value Pr(>|z|)    
## (Intercept)          0.7180159  0.0316367  22.696  < 2e-16 ***
## genderMale          -0.0350266  0.0116207  -3.014 0.002577 ** 
## age[0-10)            0.0291482  0.1679307   0.174 0.862201    
## age[10-20)           0.0457495  0.0747327   0.612 0.540422    
## age[20-30)          -0.1009230  0.0580033  -1.740 0.081867 .  
## age[30-40)          -0.0918317  0.0342574  -2.681 0.007348 ** 
## age[40-50)          -0.1039938  0.0225002  -4.622 3.80e-06 ***
## age[50-60)          -0.1341297  0.0181673  -7.383 1.55e-13 ***
## age[60-70)          -0.0803883  0.0165710  -4.851 1.23e-06 ***
## age[80-90)           0.0648350  0.0174045   3.725 0.000195 ***
## age[90-100)          0.1144732  0.0347779   3.292 0.000996 ***
## raceMissing          0.0544642  0.0386178   1.410 0.158440    
## raceAfricanAmerican  0.1155815  0.0150854   7.662 1.83e-14 ***
## raceOther            0.0673061  0.0304816   2.208 0.027238 *  
## admit_type_id2       0.1252046  0.0152659   8.202 2.37e-16 ***
## admit_type_id3      -0.0857353  0.0163343  -5.249 1.53e-07 ***
## admit_type_id4      -0.0226724  0.0200001  -1.134 0.256956    
## metforminDown        0.0682265  0.0716691   0.952 0.341114    
## metforminSteady     -0.0156810  0.0152461  -1.029 0.303703    
## metforminUp          0.1526453  0.0472482   3.231 0.001235 ** 
## insulinDown         -0.0112317  0.0186104  -0.604 0.546165    
## insulinSteady       -0.0222052  0.0137738  -1.612 0.106933    
## insulinUp            0.0236546  0.0188066   1.258 0.208470    
## readmitted<30        0.0792661  0.0189607   4.181 2.91e-05 ***
## readmitted>30        0.0422792  0.0125895   3.358 0.000784 ***
## num_procs            0.0111642  0.0036594   3.051 0.002282 ** 
## num_meds             0.0308567  0.0007081  43.577  < 2e-16 ***
## num_ip               0.0140515  0.0044002   3.193 0.001406 ** 
## num_diags            0.0273075  0.0035156   7.768 8.00e-15 ***
## ---
## Signif. codes:  0 '***' 0.001 '**' 0.01 '*' 0.05 '.' 0.1 ' ' 1
## 
## (Dispersion parameter for poisson family taken to be 1)
## 
##     Null deviance: 13093.4  on 6999  degrees of freedom
## Residual deviance:  9800.8  on 6971  degrees of freedom
## AIC: 31882
## 
## Number of Fisher Scoring iterations: 5
\end{verbatim}

\newpage
\section{Feature selection} \label{App:AppendixG}

\begin{verbatim}
## Start:  AIC=35125.1
## days ~ 1
## 
##                 Df Deviance   AIC
## + num_meds       1    10340 32381
## + num_diags      1    12539 34579
## + num_procs      1    12581 34622
## + insulin        3    12896 34954
## + age            9    12922 35033
## + readmitted     2    13015 35064
## + num_ip         1    13033 35074
## + admit_type_id  3    13028 35086
## + gender         1    13072 35113
## + metformin      3    13066 35125
## <none>                13093 35125
## + race           3    13074 35132
## 
## Step:  AIC=32380.54
## days ~ num_meds
## 
##                 Df Deviance   AIC
## + admit_type_id  3    10168 32235
## + num_diags      1    10215 32265
## + age            9    10165 32286
## + readmitted     2    10296 32355
## + num_ip         1    10316 32365
## + gender         1    10320 32369
## + race           3    10309 32376
## <none>                10340 32381
## + metformin      3    10317 32384
## + num_procs      1    10339 32389
## + insulin        3    10335 32402
## 
## Step:  AIC=32234.82
## days ~ num_meds + admit_type_id
## 
##              Df Deviance   AIC
## + num_diags   1    10071 32147
## + age         9    10004 32150
## + readmitted  2    10132 32217
## + num_ip      1    10147 32223
## + gender      1    10148 32224
## + race        3    10138 32232
## <none>             10168 32235
## + num_procs   1    10166 32242
## + metformin   3    10150 32244
## + insulin     3    10162 32256
## 
## Step:  AIC=32146.54
## days ~ num_meds + admit_type_id + num_diags
## 
##              Df Deviance   AIC
## + age         9   9937.3 32093
## + race        3  10029.5 32132
## + gender      1  10051.8 32137
## + readmitted  2  10045.5 32139
## + num_ip      1  10056.6 32141
## <none>           10070.6 32147
## + num_procs   1  10068.8 32154
## + metformin   3  10056.0 32158
## + insulin     3  10065.3 32168
## 
## Step:  AIC=32092.94
## days ~ num_meds + admit_type_id + num_diags + age
## 
##              Df Deviance   AIC
## + race        3   9872.5 32055
## + readmitted  2   9909.3 32083
## + num_ip      1   9919.4 32084
## + gender      1   9925.8 32090
## <none>            9937.3 32093
## + num_procs   1   9933.1 32098
## + metformin   3   9925.2 32107
## + insulin     3   9931.8 32114
## 
## Step:  AIC=32054.74
## days ~ num_meds + admit_type_id + num_diags + age + race
## 
##              Df Deviance   AIC
## + readmitted  2   9844.1 32044
## + num_ip      1   9857.1 32048
## <none>            9872.5 32055
## + gender      1   9865.2 32056
## + num_procs   1   9867.9 32059
## + metformin   3   9861.0 32070
## + insulin     3   9865.9 32075
## 
## Step:  AIC=32043.96
## days ~ num_meds + admit_type_id + num_diags + age + race + readmitted
## 
##             Df Deviance   AIC
## <none>           9844.1 32044
## + num_ip     1   9835.9 32045
## + gender     1   9836.6 32045
## + num_procs  1   9837.9 32047
## + metformin  3   9832.0 32059
## + insulin    3   9838.2 32065
\end{verbatim}

\begin{verbatim}
## 
## Call:  glm(formula = days ~ num_meds + admit_type_id + num_diags + age + 
##     race + readmitted, family = poisson(link = "log"), data = data.train)
## 
## Coefficients:
##         (Intercept)             num_meds       admit_type_id2  
##             0.68976              0.03196              0.13040  
##      admit_type_id3       admit_type_id4            num_diags  
##            -0.08048             -0.02164              0.02825  
##           age[0-10)           age[10-20)           age[20-30)  
##             0.01025              0.04922             -0.07774  
##          age[30-40)           age[40-50)           age[50-60)  
##            -0.08675             -0.10475             -0.13264  
##          age[60-70)           age[80-90)          age[90-100)  
##            -0.08221              0.06534              0.11422  
##         raceMissing  raceAfricanAmerican            raceOther  
##             0.04780              0.12000              0.06826  
##       readmitted<30        readmitted>30  
##             0.08637              0.04778  
## 
## Degrees of Freedom: 6999 Total (i.e. Null);  6980 Residual
## Null Deviance:       13090 
## Residual Deviance: 9844  AIC: 31910
\end{verbatim}

\section{Final Model - poisson/log model on reduced variables } \label{App:AppendixH}

\begin{verbatim}
## 
## Call:
## glm(formula = days ~ num_meds + admit_type_id + num_diags + age + 
##     race + readmitted, family = poisson(link = "log"), data = data.train)
## 
## Deviance Residuals: 
##     Min       1Q   Median       3Q      Max  
## -3.8319  -0.9928  -0.2776   0.5837   4.8197  
## 
## Coefficients:
##                       Estimate Std. Error z value Pr(>|z|)    
## (Intercept)          0.6897575  0.0306682  22.491  < 2e-16 ***
## num_meds             0.0319615  0.0006071  52.650  < 2e-16 ***
## admit_type_id2       0.1304010  0.0150795   8.648  < 2e-16 ***
## admit_type_id3      -0.0804769  0.0160805  -5.005 5.60e-07 ***
## admit_type_id4      -0.0216397  0.0199701  -1.084 0.278540    
## num_diags            0.0282470  0.0034816   8.113 4.93e-16 ***
## age[0-10)            0.0102479  0.1677619   0.061 0.951291    
## age[10-20)           0.0492239  0.0742600   0.663 0.507421    
## age[20-30)          -0.0777389  0.0574036  -1.354 0.175656    
## age[30-40)          -0.0867475  0.0340017  -2.551 0.010733 *  
## age[40-50)          -0.1047545  0.0224097  -4.675 2.95e-06 ***
## age[50-60)          -0.1326376  0.0180902  -7.332 2.27e-13 ***
## age[60-70)          -0.0822150  0.0165353  -4.972 6.62e-07 ***
## age[80-90)           0.0653429  0.0173310   3.770 0.000163 ***
## age[90-100)          0.1142206  0.0346141   3.300 0.000967 ***
## raceMissing          0.0477977  0.0385507   1.240 0.215025    
## raceAfricanAmerican  0.1200019  0.0149734   8.014 1.11e-15 ***
## raceOther            0.0682570  0.0304596   2.241 0.025033 *  
## readmitted<30        0.0863735  0.0185835   4.648 3.35e-06 ***
## readmitted>30        0.0477848  0.0124059   3.852 0.000117 ***
## ---
## Signif. codes:  0 '***' 0.001 '**' 0.01 '*' 0.05 '.' 0.1 ' ' 1
## 
## (Dispersion parameter for poisson family taken to be 1)
## 
##     Null deviance: 13093.4  on 6999  degrees of freedom
## Residual deviance:  9844.1  on 6980  degrees of freedom
## AIC: 31907
## 
## Number of Fisher Scoring iterations: 5
\end{verbatim}

\newpage
\section{q‐q plot} \label{App:AppendixI}

\includegraphics[width=0.5\linewidth]{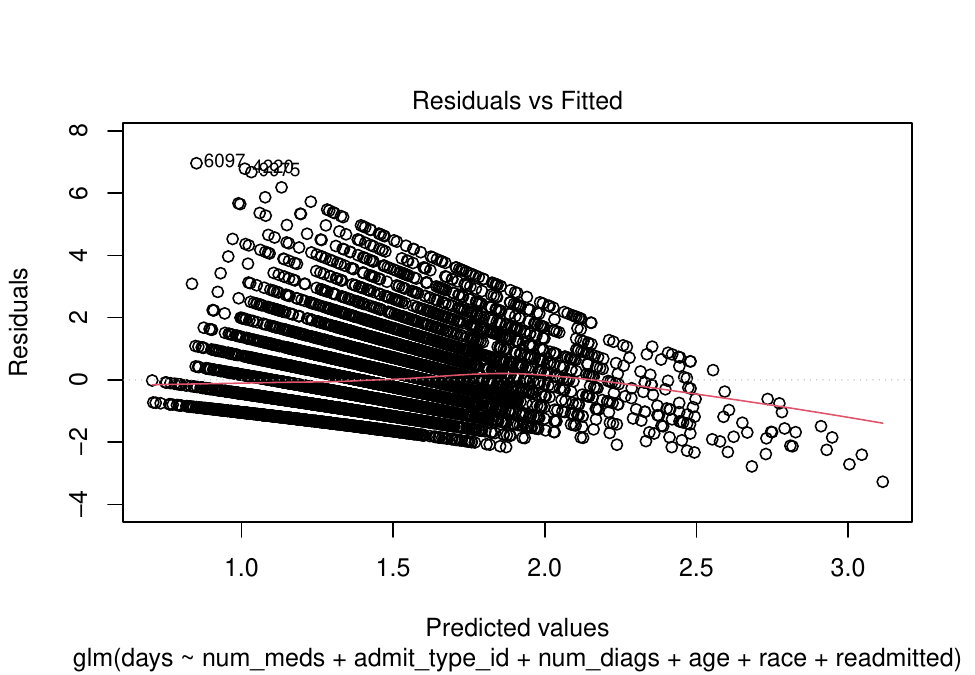} \includegraphics[width=0.5\linewidth]{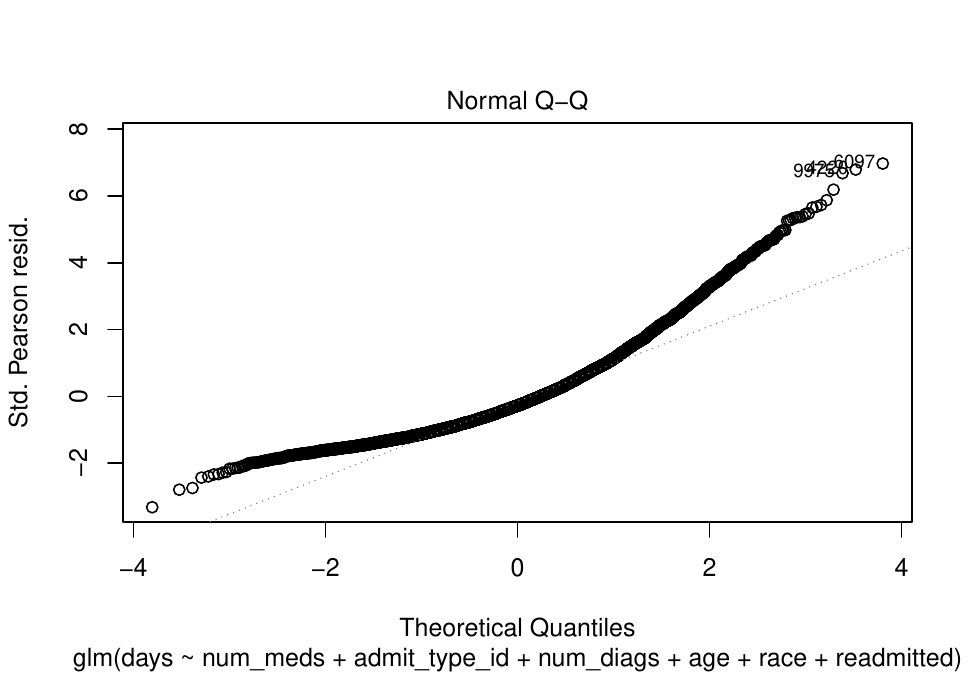} \includegraphics[width=0.5\linewidth]{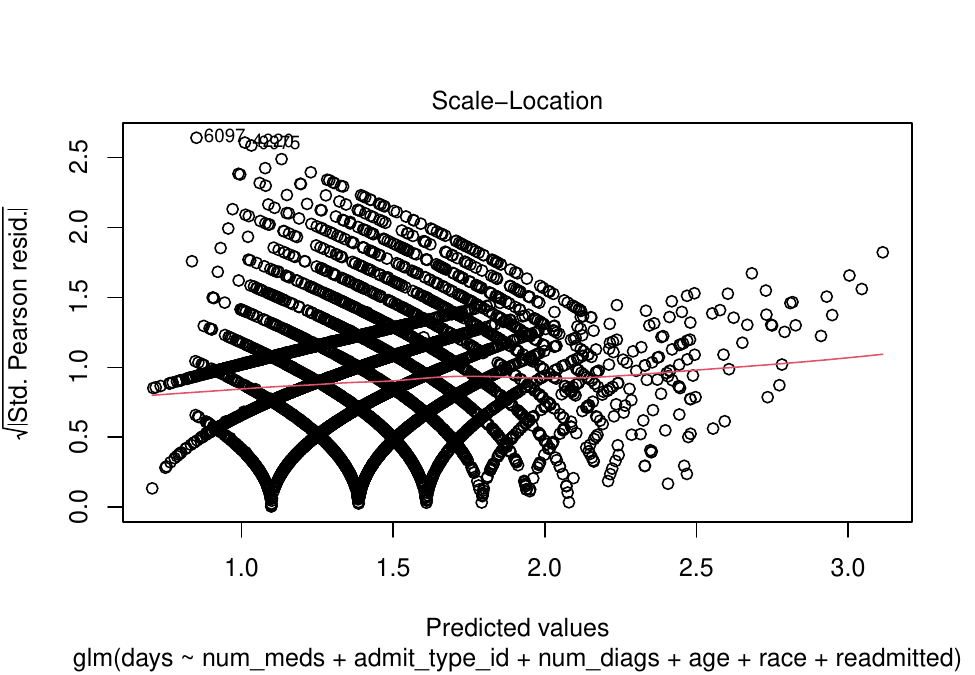} \includegraphics[width=0.5\linewidth]{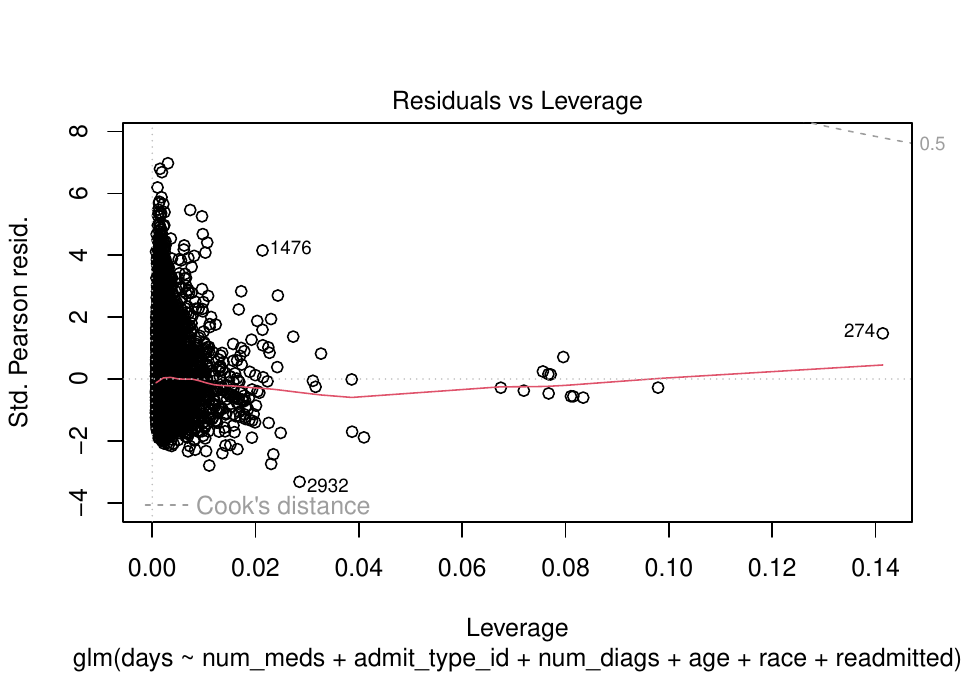}
\end{document}